\documentstyle[12pt]{article}

\begin{document}

\def\beq{\begin{equation}}
\def\eeq{\end{equation}}
\def\ds{\rm de Sitter~}
\def\fs{\rm fluctuations~}
\def\jm{j^{\mu}}
\def\fr{{<\phi^{2}>}_{R}}
\def\g5{{\gamma}_{5}}
\def\sm{$ {\rm T}_{\rm SM} $}
\def\beqa{\begin{eqnarray}}
\def\eeqa{\end{eqnarray}}
\def\b2{{\rm B}^{2}}
\def\mt{{\rm m}_{t}}
\def\m2{{\mu}^{2}}
\def\in{$\frac{1}{\rm N}$}
\newcommand{\etal}{\mbox{\it et al.}}
\newcommand{\prdj}[1]{{ \it Phys.~Rev.}~{\bf D{#1}}}
\newcommand{\prlj}[1]{{ \it Phys.~Rev.~Lett.}~{\bf {#1}}}
\newcommand{\plbj}[1]{{ \it Phys.~Lett.}~{\bf {#1B}}}
\newcommand{\npbj}[1]{{ \it Nucl.~Phys.}~{\bf B{#1}}}
\newcommand{\ptpj}[1]{{ \it Prog.~Theor.~Phys.}~{\bf {#1}}}
\newcommand{\zfpj}[1]{{ \it Zeit.f\"{u}r Physik}~{\bf C{#1}}}
\newcommand{\pk}[1]{\phi_k(#1)}
\newcommand{\pht}[1]{\phi({\vec #1},t)}
\newcommand{\be}{\begin{equation}}
\newcommand{\ee}{\end{equation}}
\newcommand{\p}{\partial}

\titlepage

\begin{flushright} IFT-15/92  \end{flushright}
\vspace{4ex}

\begin{center} \bf

GRAVITATIONALLY INDUCED SCALAR FIELD FLUCTUATIONS \\
IN THE RADIATION DOMINATED R-W UNIVERSE${}^{*}$\\
\rm
\vspace{3ex}
Zygmunt Lalak\footnote[1]{EMAIL:LALAK@FUW.EDU.PL},~
Krzysztof A. Meissner\footnote[2]{EMAIL:MEISSNER@FUW.EDU.PL},~
Jacek Pawe\l  czyk\footnote[3]{EMAIL:PAWELC@FUW.EDU.PL\\
${}^*$Work partially
supported by Polish
Government Research Grant GR-11.}~\\
Institute of Theoretical Physics \\ University of Warsaw \\
Warsaw, PL 00-681 \\

\vspace{12ex}

ABSTRACT
\\
\vspace{3ex}
\end{center}

It is shown that quantum fluctuations due to a
nontrivial gravitational background in the flat radiation
dominated universe can play an important cosmological role
generating nonvanishing cosmological global charge, e.g. baryon
number, asymmetry. The explicit form of the fluctuations at
vacuum and at finite temperature is given. Implications for
particle physics are discussed.\\

\newpage

\noindent{\rm \large 1. Introduction}
\vspace{2ex}

Since the early eighties it has been widely recognized that quantum
fluctuations of
scalar matter fields may play an important role in cosmology, especially  in
the context of the inflationary de Sitter epoch [1].
Actually, the case of the
scalar field in the de Sitter space has been the most extensively and
carefully studied one. The reason is that in the \ds space \fs of the light
fields ($m^2/H_I^2 << 1$, where $H_I$ is the \ds Hubble parameter) grow
linearly with time assuming finally a significantly large value
of the order of $\frac{{\rm H}^{4}_{\rm I}}{{\rm m}^{2}}$,which
singles out the \ds universe.
It is believed that fluctuations produced at the
time of inflation are seen during subsequent stages of the evolution
of the universe as energy density inhomogeneities
responsible for the formation of the large scale structure. It
is also advocated that those fluctuations set initial conditions
for the classical evolution of fields in subsequent epochs.

In contrast to the above,
it is usually assumed that  gravitationally induced scalar field fluctuations
in spatially flat radiation dominated (RD) and matter dominated (MD) epochs
are irrelevant for particle cosmology.
We would like to point out that this assumption
is not properly discussed in the literature. On one hand, one observes
that in the RD universe the fluctuations
(as explained in this letter) decrease in time.
On the other hand, they may in principle be large
enough to control violation of some symmetries or to alter the
evolution of some fields present in field-theoretical models. We want
to stress that this problem becomes particularly important in view of
the ongoing search for a reliable mechanism for production of the
baryon asymmetry in the Universe, the need of the better understanding
of the scenarios for late-time phase transitions and discussions
of the possible lepton number nonconservation.

In this letter we would like to  address explicitly the problem
of quantum fluctuations of the massive scalar field during the RD epoch.
This epoch covers most of the history of the universe,
and the temperature range from, say, $10^{14}$ GeV down to $10$ eV.
On that energy scale one can find a lot of interesting phenomena
in popular extensions of the standard model such as its
supersymmetric version or string inspired models, what, in our
opinion, justifies the research reported in this work.

The paper is organized as follows. In Section 1 we set our notation
and subsequently evaluate fluctuations of a massive
scalar field in the RD flat Robertson-Walker space
at vacuum and at finite temperature. In Section 2
we apply our formulae to a general field
theoretical model with particular attention paid to two specific
examples resembling the Affleck-Dine model [2] and the so called
spontaneous baryogenesis scenario [3]. Finally, in the last Section
we review our results and present conclusions.
\\
\\
\vspace{2ex}
\noindent{\rm \large 1. Scalar field fluctuations in the RD
universe\\}
\vspace{2ex}

The RD Universe is the solution to the Einstein's equations
with the energy-momentum tensor in the form
$T^\mu_\nu =diag(-\rho,p,p,p)$. Tracelessness of the
$T^{\mu}_{\nu}$  implies
equation of state for the content of the RD Universe: $\rho=3p$.
In this letter we assume a flat RD space endowed with the metric
$g_{\mu\nu}=diag(-1,a^2(t),a^2(t),a^2(t))$ where $a(t)$ is the RW
scale factor given by $a(t)\equiv (u)^{1/2}$ with $u$ defined as
$u \equiv t/t_0$,
$t_0$ being the beginning of the RD epoch.

We couple a massive scalar field to gravity in the minimal way ( note that
here the curvature scalar $R$ vanishes identically )
\be
S[\phi]=\int d^4x \sqrt{-g} \left( -g^{\mu\nu}\p_\mu\phi\p_\nu\phi
-m^2\phi^2\right).
\label{action}
\ee

As usual in this type of analysis
we assume that there is no ``back-reaction'' of the scalar field
on the metric, cf. [4].
The equation resulting from eq.(\ref{action}) is
\be
\left( \frac{d^2}{du^2} + \frac{3}{2u}\frac{d}{du} + \frac{k^2t_0^2}{u}
+m^2t_0^2\right)\phi_k(u)=0
\label{vawe}
\ee
where $\phi_k(u)$ is the spatial Fourier transform of the field
$\phi({\vec x},t)$,
\be
\phi({\vec x},t) = \int \frac{d^3k}{(2\pi)^3 2k}\left(\phi_k(u)
e^{i{\vec k}{\vec x}} a^{\dagger}_k +h.c.\right)
\label{fourier}
\ee
with $a^{\dagger}$ and its hermitian conjugate denoting standard creation and
anihilation operators respectively.

A general solution of the equation (\ref{vawe}) is given by confluent
hypergeometric functions
\begin{eqnarray}
\phi_k(u)& =&A_1(k,m) 2ikt_0 e^{-imut_0}{}_1 F_1\left(3/4+\frac{ik^2t_0}{2m},
3/2,2imt_0u\right)\nonumber\\
& +&A_2(k,m)\frac{1}{\sqrt{u}} e^{-imut_0} {}_1F_1\left(
1/4+\frac{ik^2t_0}{2m},1/2,2imt_0u\right).
\label{solution}
\end{eqnarray}
The two undetermined coefficients $A_1$ and $A_2$ , which may in principle
depend on both $m$ and $k$, are not independent if one takes into
account quantization condition imposed on a field $\phi$
\begin{eqnarray}
\left[\pht{x},\p_t\pht{y}\right]&=&
\frac{i}{\sqrt{-g}}\delta^{(3)}({\vec x}-{\vec y})\\
\label{comm}
\left[a_k,a^{\dagger}_{k'}\right]&=&(2\pi)^3 2k\delta^{(3)}({\vec k}-{\vec k'})
\label{comma}
\end{eqnarray}
{}From above equations and decomposition (\ref{fourier}) we get a normalization
condition
\be
{\rm Im}({\bar \phi_k(u)}\p_t\phi_k(u))=\frac{k}{\sqrt{-g}}
\label{cond}
\ee
which translates into the constraint on $A_1$ and $A_2$
( we confine them to be real )
\be
A_1(k,m)A_2(k,m)=1
\label{equa}
\ee
In most general case there are several ways of fixing both coefficients.
One possibility is to use initial conditions set at the timelike surface
$t=t_0$ for $\phi$ and $\p_t\phi$. This is the proper procedure if one
knows for example the explicit solution for $\phi$ in the epoch preceding
the RD one. We do not assume such a detailed knowledge, hence we use
an alternative approach instead. We demand that the ``correct'' mode functions
we choose, which will define our Fock space, should approach at short
distances ($k\to\infty$) the massless positive frequency solution,
\be
\phi_k(u)\to\frac{1}{u^{1/2}} e^{2ikt_0\sqrt{u}}
\label{limit}
\ee
In this way we obtain the asymptotic behaviour of both coefficients
\be
A_{1,2}(k,m)\to 1
\label{asym}
\ee
Here we assume that $A_1=A_2=1$, what completes the definition of our
Fock space.

We set out to calculate the fluctuations of the field $\phi$ i.e.
$<0|\phi^2|0>$.
This quantity is badly divergent and needs renormalization. We define
the renormalized  fluctuations as
 the difference between RD and the Minkowski space
fluctuations, hence the relevant object to look at is the difference
\be
<\phi^2>_R \equiv <\phi^2>-<\phi^2>_M
\label{fluc}
\ee
As we shall see,  this definition gives the finite result.
It may easily be checked that in terms of Fourier modes $\phi_k(u)$ the
renormalized fluctuations  (\ref{fluc}) are given by the formula
\be
<\phi^2>_R = \int \frac{d^3k}{(2\pi)^3} \left(\frac{1}{2k} |\phi_k(u)|^2
-\frac{1}{2u^{3/2}\sqrt{k^2/u+m^2}}\right),
\label{form}
\ee
 Using (\ref{solution}) we can write down
the explicit formula
\begin{eqnarray}
<\phi^2>_R&  =&\nonumber\\
&{}& \frac{1}{4\pi t^2}\int_0^\infty dy
\Bigg[y \left| {}_1 F_1\! \left(\frac{1}{4}
+\frac{iy^2}{2mt},\frac{1}{2},2imt\right)
+ 2iy\,{}_1 F_1\! \left(\frac{3}{4}
+\frac{iy^2}{2mt},\frac{3}{2},2imt\right)\right|^2\nonumber\\
&{}&-\frac{y^2}{\sqrt{y^2+m^2t^2}}\Bigg]
\label{fform}
\end{eqnarray}
where $y=kt/\sqrt{u}$.
Unfortunately, the above expression cannot be evaluated in its most
general form. However, it is possible to write down the systematic expansion
of the mode functions (\ref{solution}) and the integral (\ref{form}) in
terms of $mt$. Using such an expansion we will be able to discuss reliably
fluctuations in the regime of small mass and to control the passage to the
massless limit. For the region $mt>1$ we will have to rely on the
numerical calculations.

 In the case $mt<1$, the relevant expansion of modes is given by
(cf. [5])
\be
\phi_k(u)=\frac{1}{u}\sum_{n=0}^\infty\left[p_n^{(-1/2)}(2imt)j_{n-1}(2y)+
ip_n^{(1/2)}(2imt)j_n(2y)\right]\frac{1}{(2y)^{n-1}}
\label{bucha}
\ee
The $j_n$
is the n-th spherical Bessel function and coefficient $p_n^{(\mu)}$ can be
read from
\be
\sum_{n=0}^\infty p_n^{(\mu)}(z)w^n=e^{z/2({\rm coth}(2w)-1/2w)}\left(\frac{
zw}{{\rm sinh}(zw)}\right)^{1-\mu}
\label{buchb}
\ee
One easily finds that
\be
|\phi_k(u)|^2=\frac{1}{u}\left[1+(mt)^2 \left(\frac{{\rm sin}^2(2y)}{8y^4}-
\frac{1}{2y^2}\right) +O((mt)^4)\right]
\label{expa}
\ee
On the basis of the expansion we see that $<\phi^2>$
is the ultraviolet-finite quantity.
It is also infrared finite, since the modes are perfectly regular functions for
$k\to 0$ (i.e. $y\to 0$).

 In the region $mt<1$, with help of the expansion (\ref{expa}), we
get the following formula for the leading behaviour:
\be
<\phi^2>=\frac{m^2}{8\pi^2}\left(-{\rm ln}(mt)+(3/2-\gamma-{\rm ln}2)+O((mt)^2)
\right)
\label{malem}
\ee
where $\gamma$ is the Euler constant.
One should note that the fluctuations vanish as $m$ approaches zero and
grow with $m$ if we keep $mt$ constant. This agrees with the
earlier result for an exactly massless field reported in ref. [6].
The interesting feature of the
formula (\ref{malem}) is its non-analyticity in $mt$ and the
appearance of the logarithmic singularity at $t= 0$ which
 is related to the singularity
of the RD Universe at $t=0$.

In the regime $mt>1$ the numerical study we have performed shows
that the integral in (\ref{fform}), which is solely a function
of $ mt$, has an oscillating behaviour with the
period close to $\pi/m$ and amplitude
rising with increasing $mt$.
One can easily check that the integral (\ref{fform}) is converging rather
quickly, hence in a given interval of the variable $mt$
one can cut-off the integration from above at a numerically
determined value $\Lambda$. Given this observation, it is
straightforward to find an algebraic  approximation for the
integral as a function of $mt$. Using the expansion of the Kummer
hypergeometric function in terms of the modified Bessel
functions of the second kind (cf. [7])
\begin{eqnarray}
{}_1 F_1 (a,b,x) & = & e^{\frac{1}{2} x} \Gamma (b-a-\frac{1}{2}
) (\frac{1}{4} x )^{a-b+\frac{1}{2}} \nonumber \\
{}               & \times &  \sum_{n=0}^{\infty}
\frac{(2b-2a-1)_n (b-2a)_n (b-a-\frac{1}{2}
+n)}{ n! (b)_n } \nonumber \\
{} & \times & (-1)^n I_{b-a-\frac{1}{2} + n} (\frac{1}{2} x )
\end{eqnarray}
and expanding the definite
integrals in
powers of $\Lambda^2 / mt$ one obtains in the region $mt> \Lambda^2$
the expression of the form
\beq
\fr = \frac{1}{t^2} \sqrt{mt} (a \sin (2mt) + b \cos (2mt) + c +
o(\Lambda^2 / mt) )
\label{qlarg}
\eeq
where the coefficients a, b, c are
\beq
a = \frac{1}{(\sqrt{2} \pi)^3} \int_{0}^{\Lambda^2} dx e^{-2 \pi x} \,
Re(f_1 (x) f_{2}^{*} (x))
\eeq
\beq
b = \frac{1}{2 (\sqrt{2} \pi)^3} \int_{0}^{\Lambda^2} dx e^{-2 \pi x} \,
(|f_1 (x)|^2 - |f_{2}(x)|^2)
\eeq
\beq
c = \frac{1}{2 (\sqrt{2} \pi)^3} \int_{0}^{\Lambda^2} dx e^{-2 \pi x} \,
(|f_1 (x)|^2 + |f_{2}(x)|^2) - \frac{1}{3 \sqrt{2} \pi^2} \Lambda^3
\eeq
with functions $f_1,\, f_2$ defined as follows
\begin{eqnarray}
f_{1}(x) & = & \Gamma(-\frac{1}{4} - ix) \sum_{n=0}^{\infty}
\frac{(-1/2 -2ix)_{n} (1/4 -ix +n) (-2ix)_{n}}{n! (\frac{1}{2})_{n}}
\cos \pi /2 (n + 1/4 -ix) \nonumber \\
{}       & + & 4 \sqrt{x} \Gamma(\frac{1}{4} - ix)
 \sum_{n=0}^{\infty}
\frac{(1/2 -2ix)_{n} (1/4 -ix +n) (-2ix)_{n}}{n! (\frac{3}{2})_{n}}
\cos \pi /2 (n + 3/4 -ix) \nonumber \\
{}       & {}& {}
\end{eqnarray}
\begin{eqnarray}
f_{2}(x) & = & \Gamma(-\frac{1}{4} - ix) \sum_{n=0}^{\infty}
\frac{(-1/2 -2ix)_{n} (1/4 -ix +n) (-2ix)_{n}}{n! (\frac{1}{2})_{n}}
\sin \pi /2 (n + 1/4 -ix) \nonumber \\
{}       & + & 4 \sqrt{x} \Gamma(\frac{1}{4} - ix)
 \sum_{n=0}^{\infty}
\frac{(1/2 -2ix)_{n} (1/4 -ix +n) (-2ix)_{n}}{n! (\frac{3}{2})_{n}}
\sin \pi /2 (n + 3/4 -ix) \nonumber \\
{}       & {}& {}
\end{eqnarray}
where $a_n \equiv a_{n-1} (a+n-1), \; a_0 \equiv 1$.
For the reasonable value of $\Lambda = 3.5$ the least squares fit to
numerical data in the interval $(4.0, 17.0)$ gives $a=-5.9 \, 10^{-4},\;
b=4.2 \,10^{-3},\; c=1.5 \,10^{-3}$.
We note the leading dependence of the $\fr$ on $m$ in this
range of $mt$: the fluctuations grow proportionally to
the square root of the mass.

One should also note that the fluctuations $\fr$ as defined
here, see (11), (12), are not positive definite.
Remarkably, it can easily be shown, that if one perturbs the
definitions of the functions $A_{1,2} (k,m)$ allowing for the
appropriate dependence on the ratio $\frac{k^2}{m^2}$, which amounts to the
modification of our Hilbert space, then the regions of an oscillating
behaviour with negative values of $\fr$ are pushed towards
increasing values of the argument $mt$. Since in this letter we do not
discuss any specific modification of the Hilbert space beyond the
simple and most natural definitions (10) and (11), the
non-positiveness of $\fr$ demands special care when one considers
physical applications of the present result, as we do in the next
Section. The point is that a physically meaningful quantity which has
the interpretation of the dispersion squared should be strictly
positive definite. However, we do not rely on the oscillating
behaviour of $\fr$ in the discussion of applications of the
present result being only interested in the overall time dependence
$\sqrt{m} / t^{3/2}$. This leading time and mass dependence in the
region $mt > 1$ we believe to be universal, hence we just neglect the
scheme-dependent oscillating contributions in what follows.

Up to now we have been calculating curved space vacuum
expectation value of $\phi^2$. However, if we were to take into
account that the Universe is ``hot'', i.e. it is in fact in a mixed
state to which many-particle states may contribute significantly,
we should better calculate a thermal average of $\phi^2$, with
finite temperature effects included. Assuming thermal
equilibrium of the content of the Universe we have
\beq
< \phi^2 > |_{T \geq 0} = \int \frac{d^3 k}{{(2 \pi)}^3 2 k}
| \phi_{k} |^{2} (1 + 2 n_{k} )
\label{pht}
\eeq
where $\phi_{k}$ are modes given by (\ref{solution}),
(\ref{asym}), and $n_k$ is the
occupation number for the particles with the comoving momentum $k$.
As $n_k$ we take
\beq
n_k = \frac{1}{\exp ( \frac{1}{T} \sqrt{k^2 / u + m^2} ) - 1} =
\frac{1}{\exp ( \sqrt{k^2 / T_{R}^{2} + m^2 / T^2 } ) - 1}
\eeq
which is correct for sufficiently large k in view of the choice
(\ref{limit}). The (\ref{pht}) is again divergent. However, as usual in
finite-temperature calculations, it may be divided into $T = 0$
part and the temperature correction, among which only the former
is UV divergent. Hence, we can use the renormalization procedure
(\ref{fluc}) to get meaningful results even at $T > 0$
\beq
< \phi^2 > |_{renormalized, T \geq 0} \rightarrow \fr
+ < \phi^2 >^{T}_{R}
\eeq
where
\beq
< \phi^2 >^{T}_{R} = 2 \int \frac{d^3 k}{{(2 \pi)}^3 2 k}
| \phi_{k} |^{2} \frac{1}{\exp ( \sqrt{k^2 / T_{R}^{2} + m^2 / T^2 } ) - 1}
\label{phitr}
\eeq
This expression may be approximated analytically in two limiting
cases: a) $\frac{m}{T} >> 1$, and b) $\frac{m}{T} << 1$.
In the case b) one easily gets
\beq
< \phi^2 >^{T}_{R} = \frac{T^2}{12}
\label{phtsm}
\eeq
exactly as in the flat Minkowski case. In the case a) one can
see that in the region which dominates the integral, $k <
\frac{m T_{R}}{T}$, the second term in (\ref{solution}) is unimportant. Hence
we obtain
\beq
< \phi^2 >^{T}_{R} = \Gamma (5/4) \frac{6.64^{3/2}}{\pi^3}
(g_{*})^{3/4} \frac{1}{H_{I}^{2}} ( T_{R} / T )^{4}
\frac{m^{5/2}}{M_{P}^{3/2}} T^{3} e^{-m/T} \cos^{2} (mt -
\frac{3}{8} \pi )
\label{phtlm}
\eeq
which is exponentially suppressed.
 \\
 \\
\vspace{2ex}
\noindent{\rm \large 2. Implications for particle physics
in the expanding Universe\\}
\vspace{2ex}

Let us consider a global U(1) symmetry realized in a single
complex scalar field model. If Q is the charge of the field
$\chi$, the Noether current associated with that symmetry is
\beq
j^{\mu} = iQ \{ \bar{\chi} \partial^{\mu} \chi
 - \chi \partial^{\mu} \bar{\chi}\}
\eeq
(we put Q$\equiv$1 in what follows) and the
conservation law for $\jm$ in the expanding Universe reads
\beq
\partial_{\mu}(a^{3}(t)\jm) = -i a^{3}(t) \{ \bar{\chi}
\frac{\partial V}{\partial \bar{\chi}} - \chi
\frac{\partial V}{\partial \chi} \}
\label{diva}
\eeq
One can see that a symmetry is broken once the rhs of (\ref{diva})
is nonvanishing. One can see also that when a symmetry is
broken explicitly, the net cosmological charge density gets
generated according to the formula
\beq
a^{-3} \frac{d}{dt} (j^{0} a^{3}(t)) \approx  -i \{ \bar{\chi}
\frac{\partial V}{\partial \bar{\chi}} - \chi
\frac{\partial V}{\partial \chi} \}
\eeq
Let us assume that the term violating the symmetry
is
\beq
\delta V = \frac{\lambda}{2 n \Lambda^{2 n}} {\phi}^{2 n}
\eeq
where $\phi \equiv Re(\chi)$
(in this section we assume the absence of derivative couplings,
they will be discussed later). Suppose that the initial
conditions and the shape of the potential are such that the
$Im(\chi)$ and its fluctuations are negligible when compared with
$Re(\chi)$ at any time t (this situation may be easily realized
in the Affleck-Dine model, cf. [8]).
Hence
\beq
a^{-3} \frac{d}{dt} (j^{0} a^{3}(t)) \approx
 i \frac{\lambda}{\Lambda^{2 n}} {\phi}^{2 n}
\eeq
We can see that the magnitude of the symmetry violation is
proportional to a coupling $\lambda$, inverse powers of some
scale $\Lambda$ if $ n>2$, and to some power of the scalar field $\phi$.
In this sense one can say that, $\lambda$ and $\Lambda$ being
fixed in a given theory, it is the $\phi$ what determines the
amount of symmetry breaking. Here the quantum fluctuations of
the field $\phi$ come into play. In the quasiclassical picture
one can describe the evolution of the quantum field, lets call
it $\Phi$, writing it down as the superposition of the
quasiclassical field $\phi$ which obeys essentially classical
(perhaps perturbatively corrected) equation of motion and
quantum fluctuations $\delta \phi$, the dispersion squared of
which we identify as $\fr$.
If the potential drives the
quasiclassical field to zero, then it may happen that the
magnitude of the symmetry breaking term is determined by the
dispersion of $\delta \phi$. Using $ {< \phi >}^{2 n}
= {(\fr)}^{n}$ one gets an estimate
\beq
 a^{-3} \frac{d}{dt} (j^{0} a^{3}(t)) =
 i \frac{\lambda}{\Lambda^{2 n}} {(\fr)}^{n}
\eeq
Of course, whether this term is really a dominant one or not, it depends
on the relative magnitude and time-dependence of $\fr$ and the
classical  part of the field. We shall investigate this issue
later in this work.

At this point let us discuss the explicit form of $\fr$  as a
function of the temperature.
We know that the time dependence for
this quantity is given by (\ref{malem}) if $t < \frac{1}{m}$ and
by (\ref{qlarg}) if $t > \frac{1}{m}$.
Let us say assume that the RD epoch starts at $t=t_{0}$. Then
$m t \approx \frac{m}{H(t_{0})} \frac{a^{2}(t)}{a^{2}(t_{0})}$,
where $H(t_{0})$ is of the order of the Hubble parameter during
inflation, $H(t_{0}) \approx H_{I} \approx 10^{14}$ GeV.
Assuming then an adiabatic expansion in the RD epoch we get
\beq
\fr \approx \frac{m^2}{2} \ln ( \frac{H_{I} T^{2}}{m T_{R}^{2}})
\label{qsmt}
\eeq
as long as $T^2 >> T_{R}^{2} \frac{m}{H_{I}}$ where $T_{R}$ is
the reheating temperature after inflation
\beq
T_{R} = {(\frac{45}{4 \pi^3 g_{*}})}^{\frac{1}{4}} min[{(H_{I}
M_{P})}^{\frac{1}{2}},{(\Gamma_{I} M_{P})}^{\frac{1}{2}}]
\eeq
(here $g_{*}$ is the number of relativistic degrees of freedom
at of reheating and $\Gamma_{I}$ is the total decay width of
the inflaton field -- a field which drives the transition from
the \ds to RD epoch). Let us take for simplicity the case of
a ``good'' reheating which corresponds to $T_{R} \approx H_{I}$.
This gives the condition
\beq
T > T_{*} = {(m H_{I})}^{\frac{1}{2}}
\eeq
If we take $m \leq 10^{2} \, {\rm GeV}, \,H_{I} \approx 10^{14}
\, {\rm GeV}$ then we get $T_{*} \leq 10^{8} \, {\rm GeV}$. We
note that for a really soft potential with $m \approx
10^{-21}\,{\rm eV}$, one has $T_{*} \approx 10\,{eV}$ which
means that in that case the regime where $\fr$ changes only
logarithmically extends over the whole RD epoch. For $T<T_{*}$
we have instead of (\ref{qsmt})
\beq
\fr = \frac{O(1) g_{*}}{M^{2}_{P}} T^4
\eeq
($g_{*}$ is the number of relativistic degrees of freedom at
temperature T).

Now, let us investigate the evolution of the classical field $\phi$.
For simplicity we assume that this evolution is dominated by the
mass term in the potential, which is usually a good assumption
at least in the perturbative regime.
In this case the general
solution to the equation of motion is
\beq
\phi = \frac{1}{{(\frac{m t}{2})}^{\frac{1}{4}}} [C J_{1/4} (m
t) + D J_{-1/4} (m t)]
\eeq
If we set initial conditions at $t_0$ such that $z_0 = m t <<
1$, then $C = \Gamma (5/4) \phi_0 + \frac{2 \partial_t \phi_0
z_0}{m}$ and $D = - \frac{4 \Gamma (3/4) \partial_t \phi_0 {z_0}^{3/2}}
{m 2^{3/2}}$ where $\phi_0 =\phi (m t_0 ), \; \partial_t
\phi_{0} = \partial_t \phi (m t_{0})$. This gives the $\phi$ at
late times, $m t > 1$, in the form
\beq
\phi \approx \frac{1}{\sqrt{\pi}} {(\frac{m t}{2})}^{-3/4}
[\Gamma (5/4) \phi_0 + \frac{2 z_{0} \partial_t \phi_{0}}{m}]
\cos ( m t - 3 \pi / 8 )
\label{ficl}
\eeq
{}From previous analysis we have learnt that for $T < T_{*}$ the
$\fr$ falls off as $\frac{\sqrt{mt}}{t^2}$, hence we conclude that
$\frac{\fr}{\phi^2} |_{T < T_{*}} \sim \frac{m^2}{
[\Gamma (5/4) \phi_0 + \frac{2 z_{0} \partial_t \phi_{0}}{m}]^2}$
which does not depend on time (and temperature).
This implies that if the initial conditions happen to make this ratio
large, then the fluctuations dominate over the classical part of the
field.
Moreover, in
general the field $\phi$ has some additional couplings to light
particles, which facilitate decay of the field $\phi$ with the
decay width $\Gamma_{\phi}$. This changes the behaviour of the
classical field $\phi$, namely $ \phi^2 \rightarrow \exp (-
\Gamma t ) \; \phi^2 $. Actually, as pointed out by several
authors in the context of the Affleck-Dine mechanism (which
corresponds to our toy model when n=2) the $\Gamma_{\phi}$
should be large in order to avoid an unobservable excess of the
net charge produced during symmetry violation [8]. We want to
stress that in such a case, the $\fr$, decaying accordingly to
the power law, dominates the divergence of the Nother current
and the net cosmological charge density even at the late times.

Let us check whether at $T > T_{*}$ the $\fr$ may be
significantly large. From (\ref{qsmt}) and
(\ref{ficl}) we obtain, averaging
over oscillations,
\beq
\frac{\fr}{\phi^2} |_{T > T_{*}} \approx \frac{2 m^2}{{(\Gamma
(5/4) \phi_{0} + \frac{2 z_{0} \partial_t \phi_{0}}{m})}^{2}}
\eeq
That means that fluctuations are important as long as the
condition
\beq
\phi_{0} + \frac{z_{0} \partial_t \phi_{0}}{m} < m
\label{cod}
\eeq
is fulfilled. One can see that even if $m << H_{I}$ and
$\phi_{0} \sim H_{I}$, which happens to be the case if the
initial conditions at $t = t_{0}$ are produced by large quantum
fluctuations in the preceding de Sitter epoch [8], the condition
(\ref{cod}) may be fulfilled provided that $\partial_{t} \phi_{0}$ is
sufficiently large and negative.

Finally, let us consider models where a massive scalar $\phi$ is
derivatively coupled to other particle species. This situation
corresponds for instance to models possessing pseudogoldstone bosons
with nonvanishing masses. The relevant scenario is similar to
that of the
``spontaneous baryogenesis'' described in ref. [5]. If a
Lagrangian  has a coupling of the form $L_{\phi} =
- \frac{1}{f} \phi \partial_{\mu} j^{\mu}$~~(f being some,
presumably large, mass scale) where $\partial_{\mu} j^{\mu}$ is
a divergence of a current corresponding to some explicitly
broken symmetry, the baryon number symmetry for instance. Then,
as we have shown, there are fluctuations in the field $\phi$
with dispersion $\sqrt{\fr}$. We may represent them as the
effective term in the Lagrangian
\beq
L_{\delta \phi } = - \frac{1}{f} \sqrt{\fr} \partial_{\mu} j^{\mu}
\label{lef1}
\eeq
Up to the total divergence (\ref{lef1}) is equivalent to
\beq
L_{\delta \phi } =  \frac{1}{f} \partial_{0} \sqrt{\fr}  j^{0}
\eeq
This produces an effective chemical potential $\mu = -
 \frac{1}{f} \partial_{0} \sqrt{\fr}$ for the charge density $j^0$
which means a nonzero cosmological charge density generated in
thermal equilibrium. Explicitly, cf. [9],
\beq
j^{0} \approx  - \frac{1}{f} \partial_{t} \sqrt{\fr}  T^{2}
\eeq
or charge to entropy ratio
\beq
j^{0} / s  \approx  - \frac{1}{f g_{*} T } \partial_{t} \sqrt{\fr}
\eeq
where $g_{*}$ is the number of relativistic degrees of freedom
at temperature T. The above estimate gives in the case of our
toy model
\beq
j^{0} / s \approx \frac{m}{4 g_{*} f T} \frac{1}{t \sqrt{\ln (1/mt)}}
\label{cherl}
\eeq
for $T > T_{*}$ and
\beq
j^{0} / s \approx \frac{O(10^{-2})}{g_{*} f T} \frac{1}{t^{7/4}}
 m^{1/4}
\label{chers}
\eeq
for  $T < T_{*}$.
One can see that both expressions fall off as time elapses, and
that the decrease at $T < T_{*}$ is faster than at $T > T_{*}$,
essentially $\sim T^{5/2}$ below $T_{*}$ and $\sim T$ above.
If there is no phase transition in the model before the end of
RD epoch, then the final charge to entropy ratio produced will be
equal to (\ref{cherl}) or
(\ref{chers}) taken at the ``decoupling'' temperature $T_{D}$.
This is the temperature  at which
symmetry violating interactions fall off from equilibrium or the
one which corresponds to
the end of RD stage, when the shape of the fluctuations changes qualitatively
i.e. at $T_{D} \approx T_{f}$ close to 10 eV.
As previously, the numerical values predicted depend on various
details of a model under investigation. For example, let us take
$T_{D} = 10$ eV and $g_{*}=100$. Then if we require the
charge-to-entropy ratio to be equal to $10^{-10}$, as it
should be for the baryonic charge, then we get the condition
$m  =  f^4  \times 10^{-77}$GeV, which gives $m=10^{-17}$GeV for
$f=10^{15}$GeV and $m=1$GeV for
$f=10^{19}$GeV.
\\
\\
\vspace{2ex}
\noindent{\rm \large 3. Conclusions\\}
\vspace{2ex}

In this letter we have found explicit expressions for
a massive scalar field fluctuations in the flat radiation
dominated universe. It turns out that in the region of small
$mt$, i.e. shortly after the beginning of the RD epoch or
for very light fields, the fluctuations decrease with time only
logarithmically and are proportional to the square of the mass
of the field in question. For large $mt$, i.e. very late or for
a heavy field, the time dependence is stronger, $1/t^{3/2}$,
but the mass dependence becomes weaker -- proportional to the
square root of $m$. As far as finite temperatures are concerned,
 we have concluded
that the ``radiation-dominated'' background modifies Minkowski
space results rather weakly. At low temperatures, i.e. at large
ratios $m/T$, the thermal contribution is exponentially
suppressed the suppression becoming stronger as the temperature
decreases with time. At high temperatures the result coincides
essentially with that of Minkowski space. In general,
fluctuations vanish when one takes the limit $m \rightarrow 0$.

Given all that we argue that the fluctuations may still play a
significant role in particle physics models, which has been
illustrated in the second part of the work.
Within the family of models we discuss,
 the case when our parameter n equals 2 corresponds
precisely to the Affleck-Dine model, and the higher n terms are
often encountered in the important class of string inspired
models. Hence we conclude that fluctuations we have described
constitute the phenomenon which is relevant in a very general
situation when some cosmological charge density, first of all
the baryonic charge density, is supposed to be generated during
the radiation dominated epoch. We also note that
although the inflationary scenario is widely accepted, our
results do not rely on the existence of the de Sitter epoch
preceding the RD stage in the early Universe.

In conclusion,
we have demonstrated that quantum fluctuations due to a nontrivial
gravitational background during radiation dominated epoch in the
evolution of the Universe do in fact exist and may have
observable consequences for cosmology of realistic particle models.
\\
\\
\vspace{2ex}
\noindent{\large References }
\vspace{2ex}

\begin{description}

\item{[1]}
J.M.Bardeen,P.J.Steinhardt,M.S.Turner, \prdj{28}, 679 (1983);\\
A.D.Linde, \plbj{116}, 335 (1982).
\item{[2]}
I.Affleck,M.Dine, \npbj{249}, 361 (1985).
\item{[3]}
A.Cohen,D.Kaplan, \plbj{199}, 251 (1987).
\item{[4]}
N.Birrel,P.C.Davies, {\it Quantum Fields in Curved Space},
Cambridge University Press, Cambridge 1984.
\item{[5]}
H.Buchholz, {\it Die Konfluente Hypergeometrische Funktion},
Springer-Verlag, Berlin 1953.
\item{[6]}
C.Pathinayake,L.H.Ford, \prdj{37}, 2099 (1988).
\item{[7]}
M.Abramowitz,I.Stegun, {\it Handbook of Mathematical Functions},
Dover Publications, Inc., New York.
\item{[8]}
A.D.Dolgov, Yukawa Institute preprint YITP/K-940, 1991.
\item{[9]}
E.W.Kolb,M.S.Turner, {\it The Early Universe}, Addison-Wesley
Publishing Company, 1990.

\end{description}

\end{document}